\documentclass[11pt]{preprint} 
\usepackage{times}
\usepackage{mhequ}
\usepackage{mhenvs}
\usepackage{Symbols}

{\theorembodyfont{\rmfamily} 
}
\newtheorem{assumption}{Assumption}

\def\E{{\mathbf E}}
\def\P{{\mathbf P}}
\def\TV{{\mathrm{TV}}}

\begin{document}

\date{} \title{A probabilistic argument for the controllability of conservative systems}
\author{Martin Hairer}
\institute{Department of Mathematics, The University of Warwick, Coventry CV4 7AL, United
Kingdom.
\email{M.Hairer@Warwick.ac.uk}}
\titleindent=0.65cm

\maketitle \thispagestyle{empty}

\begin{abstract}
We consider controllability for divergence-free systems that have a 
conserved quantity and 
satisfy a H\"ormander condition. It is shown that such systems are  controllable, 
provided that the conserved quantity is a proper function. The proof of the result combines
analytic tools with probabilistic arguments. While this statement is well-known in geometric
control theory, the probabilistic proof given in this note seems to be new. 
We show that controllability follows from H\"ormander's condition, together with the
\textit{a priori} knowledge of  an invariant measure with full topological support for a diffusion that 
`implements' the control system.

Examples are given that illustrate the relevance of the assumptions required for the result to hold. 
Applications of the result to
ergodicity questions for systems arising from non-equilibrium statistical mechanics and to the
controllability of Galerkin approximations to the Euler equations are also given.
\end{abstract}

\section{Introduction}

In this note, we are interested in the controllability of systems of the form 
\begin{equ}[e:main]
\dot z = f(z) + u(t)\;,\quad z(t)\in \R^N\;,
\end{equ}
where $f\colon\R^N\to\R^N$ is a smooth vector field and the control $u$ is only allowed to take values 
in a given linear subspace $E$ of $\R^N$. Our aim is to study the possibility, given a starting point $z_0 \in \R^N$
and a final point $z_1 \in \R^N$, of finding a control $u$ taking values in the (possibly small) space $E$
that allows to `stir' the solution of \eref{e:main}  from $z_0$ into a small neighbourhood of $z_1$.

It is a well-known fact, see for example \cite{Lob74CON} or \cite[Chapter~3]{AS04GCT}, 
that if this control system satisfies a H\"ormander condition and  $f$ is Poisson stable, 
then \eref{e:main} is approximately controllable
 in the sense that for any two points $z_0$ and $z_1$ in $\R^N$ there
exists a time $T$ and a control $u \in \CC^\infty([0,T],E)$ such that the solution $z(T)$ of \eref{e:main} at time $T$ is equal to $z_1$. In this note, we show that this fact can be shown as a consequence
of the fact that \eref{e:main} is equivalent to a control system such that the corresponding stochastic
differential equation has the strong Feller property and possesses an invariant measure with full
topological support, see Proposition~\ref{prop:gen}. In Section~\ref{sec:nonequ}, we show how
such an argument can be reversed in order to obtain the uniqueness of the invariant measure
for an SDE where no invariant measure is known \textit{a priori}, provided that it is equivalent
in the sense of control systems to an SDE for which the invariant measure is known. This situation
arises in non-equilibrium statistical mechanics, where a non-equilibrium system can be compared
to the corresponding equilibrium system.

In the case when
$f$ is a polynomial vector field of odd degree, it is known (see for example \cite{JK85PCS} 
and references therein) that H\"ormander's condition is both necessary and sufficient for the
system \eref{e:main} to be controllable. 
This is not the case in general.
In Section~\ref{sec:counterexample}, we provide an example of a non-controllable system which
satisfies all of the conditions of the main theorem of the present article, 
except for the growth condition of the conserved quantity $H$. This illustrates the fact
 that the growth condition on $H$ encodes global geometric 
information on $f$ (Poincar\'e recurrence) that complements the local 
geometric information given by the H\"ormander condition
and is essential to the result. Applications to the controllability of the Euler equations
and to an ergodicity problem coming from non-equilibrium statistical mechanics
will be given in Sections~\ref{sec:euler} and \ref{sec:nonequ}.

\section{Setting}
\label{sec:setting}

We consider the controllability of systems that have a smooth conserved quantity called $H$, \ie we assume throughout this paper that 
\begin{equ}[e:cons]
\scal{\nabla H(z), f(z)} = 0\;,\qquad \forall z\in\R^N\;.
\end{equ}
We assume that
the state space $\R^N$ splits in a natural way as $E \oplus E^\perp$ and we denote its elements by $(x,y)$. We would like to stress the fact that even though our prime interest is Hamiltonian systems, the splitting under
consideration is not necessarily the standard splitting in position and momentum variables. Typically, $x$
would consist only of part of the momentum variables and $y$ would consist of all the other variables of the
system. In particular, we allow $\dim E \neq \dim E^\perp$, which is actually the most 
interesting case in our situation.

Our first assumption is that the flow generated by $f$ preserves the Lebesgue measure and that
it has a conserved quantity  $H$:
\begin{assumption}\label{ass:1}
The vector field $f$ is divergence-free and there exists a smooth function $H\colon\R^N\to\R$ such that \eref{e:cons} holds.
\end{assumption}

Our second assumption ensures that $H$ grows at infinity:

\begin{assumption}\label{ass:2}
The level sets $\{z\,|\,H(z) \le K\}$ are compact for every $K > 0$.
\end{assumption}

Recall that, given a vector field $f$ on $\R^N$,
we can identify it with the corresponding differential operator $\sum_i f_i(x) \d_{x_i}$. With this identification, the
Lie bracket between two vector fields is simply the vector field corresponding to the commutator of the two
differential operators. The Lie algebra generated by a family of smooth vector fields is the smallest subspace
of the space of all vector fields that is closed under the Lie bracket operation.

We also introduce an extended phase space $\R^{N+1}$, which includes time as an additional dimension,
and extend $f$ to a vector field $\tilde f$ on $\R^{N+1}$ by setting $\tilde f(x,t) = (f(x),1)$. Our last assumption
then essentially says that the differential operator $\d_t + \CL^*$ (with $\CL^*$ defined as in \eref{e:defgen} below)
is hypoelliptic, see \cite{Hor85}.

\begin{assumption}\label{ass:3}
Given a basis $\{e_1,\ldots,e_n\}$ of $E$, the Lie algebra generated by $\{\tilde f,e_i,\ldots,e_n\}$
spans $\R^{N+1}$ at every point.
\end{assumption}

Note that the statement of Assumption~\ref{ass:3} is actually independent of the particular choice of the
basis of $E$. With these notations, the main result of this article is:

\begin{theorem}\label{theo:main}
Under assumptions \ref{ass:1}--\ref{ass:3}, for every initial condition
$z_0 \in \R^N$ and every terminal condition $z_1 \in \R^N$ there exists a time $T$ and a 
control $u \in \CC^\infty([0,T],E)$
such that the solution $z(T)$ of \eref{e:main} at time $T$ is equal to $z_1$.
\end{theorem}

\section{Proof of the main result}

The main idea in the proof of \theo{theo:main} is to consider the following It\^o 
stochastic differential equation
on $\R^n$:
\begin{equs}
d\xi(t) &= f_x(\xi,\eta)\,dt - \Bigl(3 g'(H) e^{-2g(H)}\nabla_x H \Bigr)(\xi,\eta)\,dt + \sqrt 2 e^{-g(H(\xi,\eta))}\,dw(t)\;,\\
d\eta(t) &= f_y(\xi,\eta)\,dt\;,\label{e:sde}
\end{equs}
where $g\colon\R_+ \to \R$ is a function to be determined later. Here, $w$ is an $n$-dimensional standard
Wiener process.

We will assume from now on that the function $g$ is smooth, increasing, and that $g(0) = 0$.
We will also assume that $g$ grows sufficiently fast so that $\exp(-g\circ H)$ is integrable and we
denote by $Z$ the value of the integral. This can always be done thanks to Assumption~\ref{ass:2}. The following result is elementary:

\begin{lemma}
Under the above assumptions, there exists a choice of function $g$ such that \eref{e:sde} has a unique global strong solution for all times and such that
$\mu_H(dx,dy) = Z^{-1} \exp \bigl(-(g\circ H)(x,y)\bigr)\,dx\,dy$ is an invariant probability measure for \eref{e:sde}.
\end{lemma}
\begin{proof}
The existence of a unique local strong solution is a standard result for SDEs with smooth coefficients
\cite{Oks03SDE}.
To show that this solution can be continued for all times, we show that there exists a suitable choice of $g$ such
that $H$, evaluated at the solution to \eref{e:sde}, grows at most exponentially fast. Using for any function $h$
on $\R^N$ the shortcut $h_s = h(\xi(s),\eta(s))$, It\^o's formula yields
\begin{equs}
H_t &= H_0 + \int_0^t e^{-2g(H_s)}\bigl((\Delta_x H)_s - 3 \|(\nabla_x H)_s\|^2 g'(H_s)\bigr)\,ds \\
&\quad + \sqrt 2\int_0^t e^{-g(H_s)}\scal{(\nabla_x H)_s, dw(s)}\;.
\end{equs}
Since $H$ is smooth and proper, we can now chose for $g$ an increasing function that tends to $+\infty$ sufficiently fast so that
$e^{-2g(H(x,y))}(\Delta_x H)(x,y) \le C + H(x,y)$ for some constant $C$ and for every $(x,y)\in \R^N$.
Using Gronwall's inequality, this ensures that
\begin{equ}[e:apriori]
\E H_t \le H_0 \, e^{t} + C \bigl(e^{t} - 1\bigr)\;,
\end{equ}
for every $t \ge 0$. 

That $\mu_H$ is an invariant probability measure for \eref{e:sde} follows immediately from the fact that
\begin{equ}
\CL^* \exp ({-g\circ H}) = 0\;,
\end{equ}
where $\CL^*$ is the adjoint of the generator of the semigroup generated by \eref{e:sde},
\begin{equ}[e:defgen]
\CL^* F = - \nabla \cdot (f F) + \nabla_x \bigl(\nabla_x H (g'\circ H) e^{-2g\circ H} F\bigr) + \nabla_x \bigl(e^{-2g\circ H}\nabla_x F\bigr)\;.
\end{equ}
See \eg \cite{Has80SSD,Oks03SDE,RevYor99} for more details.
\end{proof}

Note furthermore that the hypoellipticity assumption~\ref{ass:3} implies that the transition probabilities
corresponding to the solutions of \eref{e:sde} have a density with respect to Lebesgue measure
that is smooth in all of its arguments (including time). This is an immediate consequence of the fact that
the transition probabilities are a solution (in the sense of distributions) to the equation
\begin{equ}
(\d_t + \CL^*)p_t(z,z') = 0\;,
\end{equ}
and that $\d_t + \CL^*$ is hypoelliptic by H\"ormander's theorem (see \cite{Hor85}).

The result stated in \theo{theo:main} is now an almost immediate conclusion of the following two facts:
\begin{claim}
\item[1.] The measure $\mu_H$ satisfies $\mu_H(A) > 0$ for every open set $A \subset \R^N$.
\item[2.] The measure $\mu_H$ is the \textit{only} invariant probability measure for \eref{e:sde} and is therefore ergodic.
\end{claim}
While the first claim is obvious, the second claim requires some more explanation. Surprisingly, it will 
turn out to be an almost immediate consequence of the first claim, once we realise that the hypoellipticity
assumption~\ref{ass:3} implies the following.

\begin{lemma}\label{lem:contTV}
Fix a time $t > 0$ and denote by $\CP_t(z,\cdot\,)$ the transition probabilities corresponding to \eref{e:sde}.
Then, they are continuous in the total variation topology. In particular, for every $z \in \R^N$, 
there exists $\delta_z > 0$ such that
\begin{equ}
\|\CP_t(z,\cdot\,) - \CP_t(z',\cdot\,)\|_\TV < 1\;,
\end{equ}
for every $z'$ such that $|z'-z| < \delta_z$. Here, $\|\,\cdot\,\|_\TV$ denotes the total variation distance between
probability measures (normalised in such a way that the distance between two mutually singular probability measures
is $2$).
\end{lemma}

\begin{proof}
This is an immediate consequence of the smoothness of the transition probabilities.
\end{proof}

Recall that the topological support $\supp \mu$ of a probability measure $\mu$ is the
smallest closed set of full $\mu$-measure. Equivalently, it is characterised as the set of points $z$ such that
every neighbourhood of $z$ has positive $\mu$-measure. As an immediate consequence of \lem{lem:contTV} 
and the fact that distinct ergodic invariant measures are mutually singular we have

\begin{corollary}\label{cor:disj}
For every $z \in \R^N$ there exists $\delta_z$ such that at most one ergodic invariant probability measure $\mu_z$ for \eref{e:sde} that satisfies $\supp \mu_z \cap \CB(z,\delta_z) \neq \emptyset$. 
Here $\CB(z,\delta)$ denotes the open ball of radius $\delta$ centred in $z$.
\end{corollary}

It follows from \cor{cor:disj} that the set of all 
ergodic invariant measures for \eref{e:sde} is countable. Denote this set by $\{\mu_i\}_{i \ge 0}$ and define
$S_i = \supp \mu_i$. An important property of the $S_i$'s which follows immediately from 
\cor{cor:disj}  is

\begin{corollary}\label{cor:finite}
Every compact region of $\R^N$ intersects at most finitely many of the $S_i$'s.
\end{corollary}

Since every invariant measure for \eref{e:sde} is a convex combination of ergodic invariant
measures, there exist weights $p_i$ with $p_i \ge 0$ and $\sum p_i = 1$ such that $\mu_H = \sum_i p_i \mu_i$.
It follows from \cor{cor:finite} that
\begin{equ}
\supp \mu_H = \bigcup \bigl\{S_i\,|\, p_i \neq 0\bigr\}\;.
\end{equ}
Since the $S_i$'s are disjoint closed sets satisfying \cor{cor:finite}, the only way in which they can cover
$\R^N$ is by having only one ergodic invariant measure with support $\R^N$. We have thus shown that
$\mu_H$ is the only invariant probability measure for \eref{e:sde} and as a consequence is ergodic.

Let us now  turn to the
\begin{proof}[of \theo{theo:main}]
Fix an arbitrary open set $A \subset \R^N$  and denote by $B$ the set of all points $z_0$ in $\R^N$ such that there
exists a time $T$ and a smooth control $u \in \CC^\infty([0,T],E)$ such that the solution of \eref{e:main} with
initial condition $z_0$ satisfies $z(T) \in A$. Our aim is to show that $B = \R^N$, which then implies
the statement of the theorem by \cite[Theorem~3.2]{Jur97}.

Consider now the solution of \eref{e:sde} with initial condition $(\xi_0, \eta_0) = z \in \R^N$ and define
the stopping time $T_z = \inf \{t> 0\,|\,(\xi(t),\eta(t)) \in A\}$. It follows immediately from Birkhoff's ergodic theorem
and the fact that $\mu_H(A) > 0$ that the set
\begin{equ}
B_0 = \{z \in \R^N\,|\, \P(T_z < \infty) = 1\}
\end{equ}
satisfies $\mu_H(B_0) = 1$, so that $B_0$ is dense in $\R^N$. Furthermore, a consequence of \lem{lem:contTV} is
that if $\CB(z,\delta_z) \cap B_0 \neq \emptyset$ then $\P(T_z < \infty) \ge \textstyle{1\over 2}$. Combining these
two statements shows that for every $z \in \R^N$ there exists a time $t \ge 0$ such that
$\CP_t(z,A) > 0$.  The support theorem \cite{StrVar72OTS}, combined with the fact that the control problem associated
to \eref{e:sde} is equivalent to \eref{e:main} (since $\nabla_x H$ takes values in $E$) allows us to conclude that $B = \R^N$.
\end{proof}

Retracing the argument of the proof, one sees that we have actually proven the following 
weaker fact:

\begin{proposition}\label{prop:gen}
Consider a control system of the form
\begin{equ}[e:maingen]
\dot z = f(z) + A(z) \, u(t)\;, 
\end{equ}
where $z \in \R^n$, $u$ is a smooth control taking values in $\R^d$, and 
$A \colon \R^n \to \R^{n\times d}$. If the corresponding stochastic differential equation
\begin{equ}
dz = f(z)\, dt + A(z)\, dw(t)\;,
\end{equ}
has the strong Feller property and possesses an invariant measure with full topological support,
then \eref{e:maingen} is approximately controllable.
\end{proposition}

\section{Examples and applications}

In this section, we present a number of examples which explore to which extent the conditions
formulated in this paper are necessary to the result. We also present a few applications in which our 
result may prove to be useful.

\subsection{Dropping the hypoellipticity assumption}

It is clear from  \cite{JK85PCS} that Assumption~\ref{ass:3} is crucial for any result of the type of \theo{theo:main},
as can be seen from the following very easy example. Consider the Hamiltonian
\begin{equ}
H = {p_1^2 + p_2^2 \over 2} + {q_1^2 + q_2^2 \over 2}\;,
\end{equ}
and define $f$ as the corresponding Hamiltonian vector field. If we define $E$ to be the
linear subspace of $\R^4$ corresponding to the variable $p_1$, it is clear that all of our assumptions
are satisfied, except for Assumption~\ref{ass:3}. However, $q_2^2 + p_2^2$ is an integral of
motion of this system that cannot be perturbed by acting on the variable $p_1$, so that the
conclusion of \theo{theo:main} does not hold.

\subsection{Dropping the conservative structure  or the growth condition on $H$}
\label{sec:counterexample}

Consider the control system in $\R^2$ given by
\begin{equ}[e:counter1]
\dot x = u(t) - x\;,\qquad \dot y = g(y+g(x)) - y \phi(y)\;,
\end{equ}
where $g\colon\R \to [-1,1]$ is an odd function with $g'(x) > 0$ for all $x$ and such that
$\lim_{x\to \pm\infty}g(x) = \pm 1$. The function $\phi\colon\R\to[0,1]$ is a smooth
function such that $\phi(y) = 0$ for $|y| \le 2$ and $\phi(y) = 1$ for $|y| \ge 3$.

One can see immediately that, whatever the values of $u$ and $x$ are, one
has $\dot y > 0$ for $y \in (1,2)$ and $\dot y < 0$ for $y \in (-2,-1)$. This shows that
the conclusions of \theo{theo:main} cannot hold in this situation. However, the
system \eref{e:counter1} is hypoelliptic since the commutator between
$\d_x$ and $-x\d_x + g(y+g(x))\d_y - y \phi(y)\d_y$ is given by
\begin{equ}
-\d_x + g'(y+g(x))g'(x)\d_y\;.
\end{equ}
This vector field always has a non-zero component in the $y$-direction because of the
assumption that $g'$ is strictly positive.

Furthermore, the orbits of \eref{e:counter1} in the absence of control stay bounded for all times
and the corresponding diffusion process has global strong solutions and 
a smooth invariant probability measure. (The above arguments actually show that it has at least two
distinct ergodic smooth invariant probability measures.) The missing point in this argument of course is
the fact that we have no explicit expression for any of these invariant measures and therefore no
\textit{a priori} knowledge about their support.

One may argue on the other hand that it is possible to find a Hamiltonian function $H(x,y)$ such that
$\d_x H(x,y) = g(y+g(x)) - y \phi(y)$ and that the control system
\begin{equs}
\dot x = -\d_y H(x,y) + u(t)\;,\qquad \dot y = \d_x H(x,y)\;,
\end{equs}
is equivalent (from the point of view taken in this paper) to \eref{e:counter1}, so that 
Assumption~\ref{ass:1} and Assumption~\ref{ass:3}
are satisfied. This example shows that the condition that the level sets of $H$ are compact
is essential for \theo{theo:main} to hold and not just a technical condition that ensures that
$\exp(-g \circ H)$ can be made integrable.

\subsection{Controllability in finite time}

Let us show by a simple example that, unlike in the polynomial case studied in \cite{JK85PCS}, it is not reasonable in general, under the conditions of
Section~\ref{sec:setting}, to expect the existence of a time $T$ independent of $z$ and $A$ such
that \eref{e:main} can be driven from $z$ to $A$ in time $T$.  Consider the function $H(x,y) = \sqrt{1 + x^2 + y^2}$
and the control system
\begin{equ}[e:counterslow]
\dot x = -\d_y H(x,y) + u(t)\;,\qquad \dot y = \d_x H(x,y)\;.
\end{equ}
It is a straightforward exercise to check that all the conditions from the previous section are satisfied, so
that \theo{theo:main} applies.

It is equally straightforward to check that $|\d_x H(x,y)| \le 1$ for every value of $x$ and $y$. Therefore, 
whatever control $u$ is used to stir \eref{e:counterslow} from $z = (x,y)$ into $A$, it will always  require
a time $T$ larger than $\inf_{(x',y') \in A} |y-y'|$.

\subsection{The Euler equations}
\label{sec:euler}

The three-dimensional Euler equations on the torus are given by
\begin{equ}[e:Euler]
\dot u(x,t) = - \bigl(u(x,t) \cdot \nabla_x\bigr) u(x,t) - \nabla p(x,t) \;,\qquad \div u(x,t) = 0\;,
\end{equ} 
where $x \in \mathbf{T}^3$. (Note that the algebraic condition on the divergence of $u$ determines
$p$ in a unique way.) If we assume that $\int u_0(x)\,dx = 0$ and expand this equation in Fourier modes, we obtain
\begin{equs}
\dot u_k = - i\sum_{h,\ell \in \Z^3\setminus\{0\} \atop h+\ell = k} (k \cdot u_h) \Bigl(u_\ell - {k\cdot u_\ell \over |k|^2} k\Bigr)
\end{equs}
subject to the algebraic conditions $k \cdot u_k = 0$ and $u_{-k} = \bar u_k$. Here, the index $k$ takes
values in $\Z^3\setminus\{0\}$ (which reflects the fact that $\int u(x,t)\,dx = 0$ for all times) and $u_k \in \R^3$.
Furthermore, the dot $\cdot$ denotes the usual scalar product in $\R^3$.

Since in this note we are only interested in finite-dimensional systems, we fix a (large) value
$N_\star$ and impose $u_k = 0$ for every $k$ such that one of its components is larger
than $N_\star$ in absolute value. 

It is a straightforward exercise to check that the total energy $\sum |u_k|^2$ is a conserved quantity for this
system and that the right-hand side of \eref{e:Euler} is divergence-free. This allows to recover immediately a weak form of the controllability results obtained in
\cite[Section~6]{Rom04EFD} and \cite{AS05NSE}. The same argument allows to show that the finite-dimensional Galerkin
approximations for the two-dimensional Euler equations are controllable under the conditions presented
in \cite{HaiMat04ENS}.

\subsection{Non-equilibrium statistical mechanics}
\label{sec:nonequ}

The articles \cite{EckPilRey99NES,EckPilRey99EPI,EckHai00NES} considered a mechanical system coupled to heat baths at different temperatures
$T_i$. This situation can be modelled by the following system of SDEs:
\begin{equs}[2]
	dq_j &= \d_{p_j}H_S\,dt - \sum_{i=1}^M \bigl(\d_{p_j}F_i\bigr) r_i\,dt\;,
&\qquad j&=1,\ldots,N\;,\\
	dp_j &= -\d_{q_j}H_S\,dt + \sum_{i=1}^M \bigl(\d_{q_j}F_i\bigr) r_i\,dt\;,
&&\label{e:chain}\\
	dr_i &= -\gamma_i r_i\,dt + \gamma_i \lambda_i^2 F_i(p,q)\,dt -
\sqrt{2 \gamma_i T_i}\,dw_i(t)\;, &\qquad i&=1,\ldots,M\;,
\end{equs}
The interpretation of this equation is that a Hamiltonian system with $N$ degrees of freedom
described by $H_S(p,q)$ is coupled to $M$ heat baths with internal states $r_i$ that are maintained
at temperatures $T_i$. The functions $F_i(p,q)$ and the constants $\gamma_i$ and $\lambda_i$ describe 
coupling between the Hamiltonian system and the $i$th heat bath, as well as the relaxation times of the heat baths.

The control problem corresponding to this system is given by
\begin{equs}[2]
	\dot q_j &= \d_{p_j}H_S - \sum_{i=1}^M \bigl(\d_{p_j}F_i\bigr) r_i\;,
&\qquad j&=1,\ldots,N\;,\\
	\dot p_j &= -\d_{q_j}H_S + \sum_{i=1}^M \bigl(\d_{q_j}F_i\bigr) r_i\;,
&&\\
	\dot r_i &= u_i(t)\;, &\qquad i&=1,\ldots,M\;,
\end{equs}
which is of the form \eref{e:main} with conserved quantity
\begin{equ}
H(p,q,r) = H_S(p,q) + \sum_{i=1}^M \Bigl({r_i^2 \over 2\lambda_i^2} - r_i F_i(p,q)\Bigr)\;.
\end{equ}
(Note that one could actually add any function of $r$ to $H$ and it would still be a conserved quantity.)
Provided that the level sets of $H_S$ are compact, it is easy to check that all of the assumptions
of Section~\ref{sec:setting} are satisfied, except for Assumption~\ref{ass:3} which has to be checked
on a case by case basis. It can for example be checked for the chain of anharmonic oscillators 
studied in the abovementioned works, provided that the nearest-neighbour coupling potential does not 
contain any infinitely degenerate point.

Note that if all the $T_i$ appearing in \eref{e:chain} are equal, then this equation is similar to \eref{e:sde}
and one can check that $e^{-H(p,q,r)/T}\,dp\,dq\,dr$ is its invariant probability measure. However, 
the mere \textit{existence} of an invariant probability measure for \eref{e:chain} with arbitrary temperatures
is an open problem in general. (In the case of a chain of oscillators it was shown in \cite{EckPilRey99NES,EckHai00NES, ReyTho02ECT} to exist, provided that the nearest-neighbour interaction
dominates the on-site potential at high energies.)

The controllability result shown in this article, combined with the support theorem \cite{StrVar72OTS} 
and the smoothness of transition probabilities for \eref{e:chain} immediately implies the following:

\begin{theorem}
If \eref{e:chain} satisfies Assumption~\ref{ass:3} and the level sets of $H_S$ are compact,
then \eref{e:chain} can have at most one invariant probability measure.
\end{theorem}

\begin{proof}
The conditions of \theo{theo:main} are satisfied by assumption (note that one can always add to $H$ a function of $r$ that grows sufficiently fast at infinity, thus ensuring that Assumption~\ref{ass:2} holds). This immediately implies that every
invariant probability measure for \eref{e:chain} has the whole phase space as its support. The
fact that every invariant measure has a smooth density allows to conclude  by the same argument
used to show that the measure $\mu_H$ is the only invariant measure for \eref{e:sde}.
\end{proof}

\begin{acknowledge}
The author would like to thank Xue-Mei Li, Luc Rey-Bellet, and Jochen Vo\ss\ for interesting
discussions on this subject.
\end{acknowledge}

\bibliographystyle{../Martin}
\markboth{\sc \refname}{\sc \refname}
\bibliography{./refs}

\end{document}